\documentclass[aps,prb,twocolumn,superscriptaddress]{revtex4}
\usepackage{amsmath}
\usepackage{graphicx}
\usepackage{esint}
\usepackage{epstopdf}
\graphicspath{{pict/}{./}}
\usepackage{bm}%

\newcounter{Fig}

\newcommand{\be}{\begin{equation}}
\newcommand{\ee}{\end{equation}}

\begin{document}

\title{Multipolar interference effects in nanophotonics}
\author{Wei Liu}
\email{wei.liu.pku@gmail.com}
\affiliation{College of Optoelectronic Science and Engineering, National University of Defense
Technology, Changsha, Hunan 410073, P. R. China}
\author{Yuri S. Kivshar}
\affiliation{Nonlinear Physics Centre,  Research
School of Physics and Engineering, Australian National University,
Canberra, ACT 0200, Australia}


\begin{abstract}
Scattering of electromagnetic waves by an arbitrary nanoscale object can be characterized by a multipole decomposition of the electromagnetic field that
allows to describe the scattering intensity and radiation pattern through interferences of dominating excited multipole modes. In modern nanophotonics,
both generation and interference of multipole modes start to play an indispensable role, and they enable nanoscale manipulation of light with many related applications.
Here we review the multipolar interference effects in metallic, metal-dielectric, and dielectric nanostructures, and suggest a comprehensive view on many phenomena
involving the interferences of electric, magnetic and toroidal multipoles, which drive a number of recently discussed effects in nanophotonics such as unidirectional
scattering, effective optical antiferromagnetism, generalized Kerker scattering with controlled angular patterns, generalized Brewster angle, and nonradiating optical anapoles. We further discuss other types of possible multipolar interference effects not yet exploited in literature and envisage the prospect of achieving more flexible and advanced nanoscale control of light relying on the concepts of multipolar interference through full phase and amplitude engineering.
\end{abstract}
\maketitle


\section{Introduction}

Stimulated by new concepts in the physics of metamaterials and metasurfaces and the rapid development of nanoscale fabrication technologies, the field of nanophotonics has experienced an explosive growth in recent years, which enables various applications relying on flexible and efficient subwavelength light manipulations (see, e.g., Refs.~\cite{NOVOTNY_principles_2012,KABASHIN_NatMater_plasmonic_2009,atwater_plasmonics_2010-1,lukyanchuk_fano_2010-1,miroshnichenko_fano_2010,NOVOTNY_Nat.Photonics_antennas_2011,
soukoulis_achievements_2011,ZHELUDEV_metamaterials_2012,KILDISHEV_Science_planar_2013,KRASNOK_Phys.-Uspekhi_optical_2013,liuwei_control_2014,BLIOKH_Nat.Photonics_spinorbit_2015,
jahani_alldielectric_2016-1,Kuznetsov_Science_2016,CHEN_Rep.Prog.Phys._review_2016}). Similar to conventional photonics operating at other spatial scales, the fundamental research and applications in nanophotonics rely significantly on resonant light-matter interactions, where the detailed analysis of the excitation and interference of electromagnetic multipoles are usually crucial and indispensable~\cite{jackson_classical_1962,BOHREN_absorption_2008,soukoulis_achievements_2011,ZHELUDEV_metamaterials_2012,PAPASIMAKIS_NatMater_electromagnetic_2016}. Conventional multipole expansions have mostly been conducted for dynamic charge-current distributions on  length scales that are much smaller than the effective wavelength of light, where consequently only the electric and magnetic dipoles are dominant~\cite{jackson_classical_1962,liu_efficient_2015,PAPASIMAKIS_NatMater_electromagnetic_2016}. Nevertheless, when the dynamic charge-current is distributed within an area comparable to or larger than the effective wavelength of light, higher order multipoles including dynamic toroidal multipoles will arise, and they would contribute to generate the electromagnetic fields~\cite{jackson_classical_1962,kaelberer_toroidal_2010-1,liu_efficient_2015,PAPASIMAKIS_NatMater_electromagnetic_2016}.  To characterize accurately the photonic features and to optimize a design of nanostructures for achieving desired functionalities, comprehensive investigations should be conducted into the physics of interference between all three families of dynamic electromagnetic multipoles, which may open an extra dimension for the nanoscale light manipulation incubating more in-depth fundamental research and realistic applications in nanophotonics.

\begin{figure}
\centerline{\includegraphics[width=8cm]{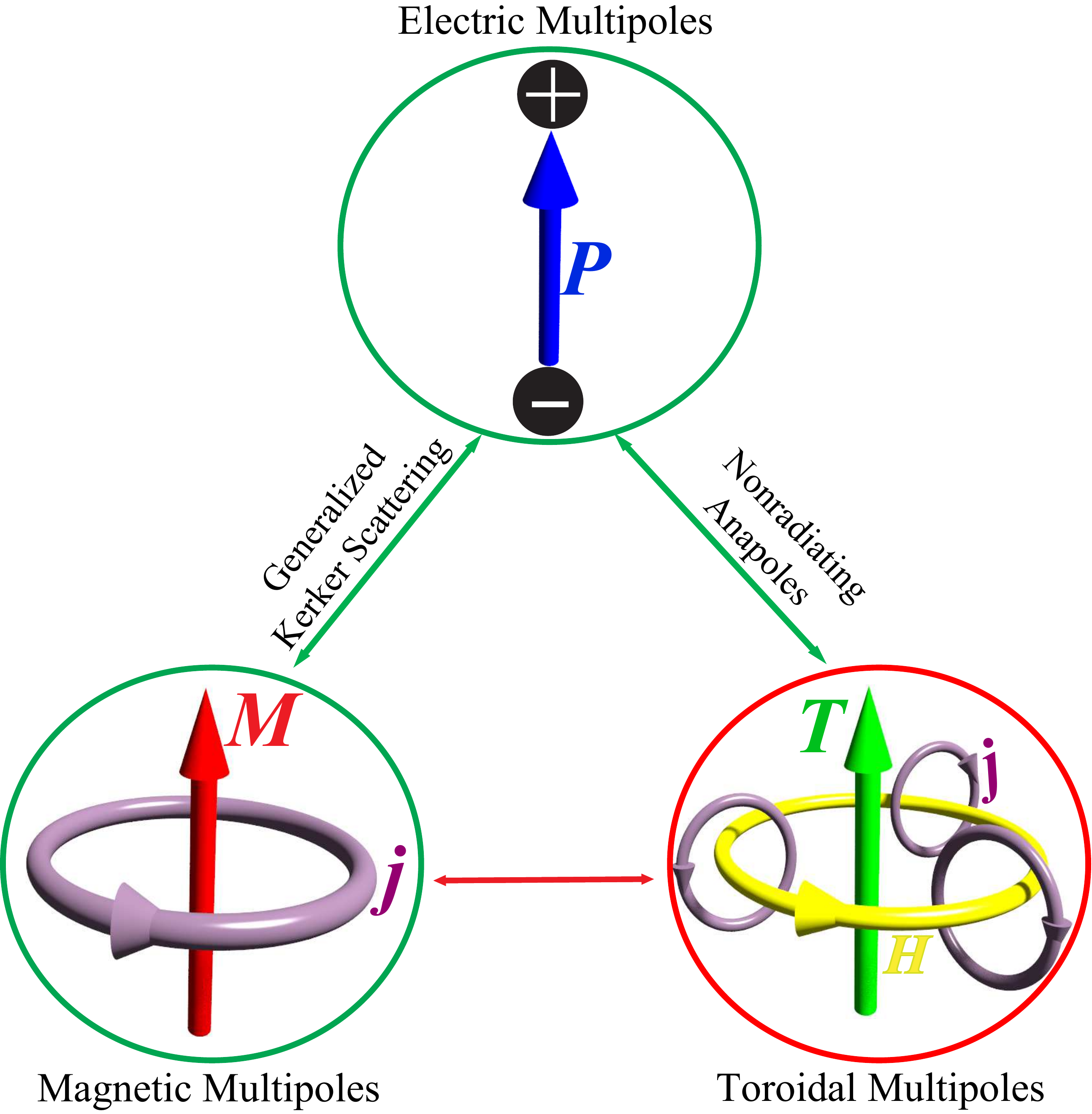}}\caption{\textbf{Schematic of interference effects discussed in this review}. We consider interferences for three families of electromagnetic multipoles: namely, electric, magnetic and toroidal multipoles. The circles denote interferences between multipoles from the same family while the double arrows stand for interferences between multipoles from different families. The corresponding charge-current distributions for the dipoles of each family are shown within the circles ($\mathbf{P}$: ED moment; $\mathbf{M}$: MD moment; $\mathbf{T}$: TD moment; $\mathbf{j}$: electric current; $\mathbf{H}$: magnetic field). The green (red) colour denotes the effects that have (have not) been studied in the existing literatures in the field of nanophotonics. (Online version in colour.)}
\label{fig1}
\end{figure}

In this review, for the first time to our knowledge, we present a brief while coherent view on different types of interference effects that may occur for the three families of classical dynamic multipole radiation modes, namely {\em electric, magnetic and toroidal multipoles}.  We discuss how such interferences enable various recently predicted or observed novel effects with nanoscale structures guiding the manipulation of light and related functionalities at the nanoscale.  The corresponding charge-current distributions for the lowest-order dipoles, such as electric dipole (ED), magnetic dipole (MD) and toroidal dipole (TD),  are shown schematically in Fig.~\ref{fig1}.  The interference effects we discuss here  can roughly be divided into two categories: (i) interferences between multipoles from the same family (circles) and (ii) interferences between multipoles from different families (double arrows).  In Sec.~\ref{ee} and Sec.~\ref{mm},  we review respectively the interferences between electric multipoles and those between magnetic multipoles. The interferences between electric and magnetic multipoles are discussed in Sec.~\ref{em}, where we focus mainly on the generalized Kerker scattering and  other induced effects such as the generalized Brewster angle and broadband high-efficient light control with metasurfaces. In Sec.~\ref{et}, we summarize the recent studies of interference effects involving electric and toroidal multipoles, also highlighting the concept of {\em nonradiating anapoles} realized with simple individual nanoparticles. Other types of possible interferences include the interferences between toroidal multipoles and interferences between magnetic and toroidal multipoles, as presented by the red double arrow and red circle in Fig.~\ref{fig1}, are discussed briefly in Sec.~\ref{co}
that concludes the paper.

 \section{Electric multipoles}
 \label{ee}

\begin{figure*}
\centerline{\includegraphics[width=13cm]{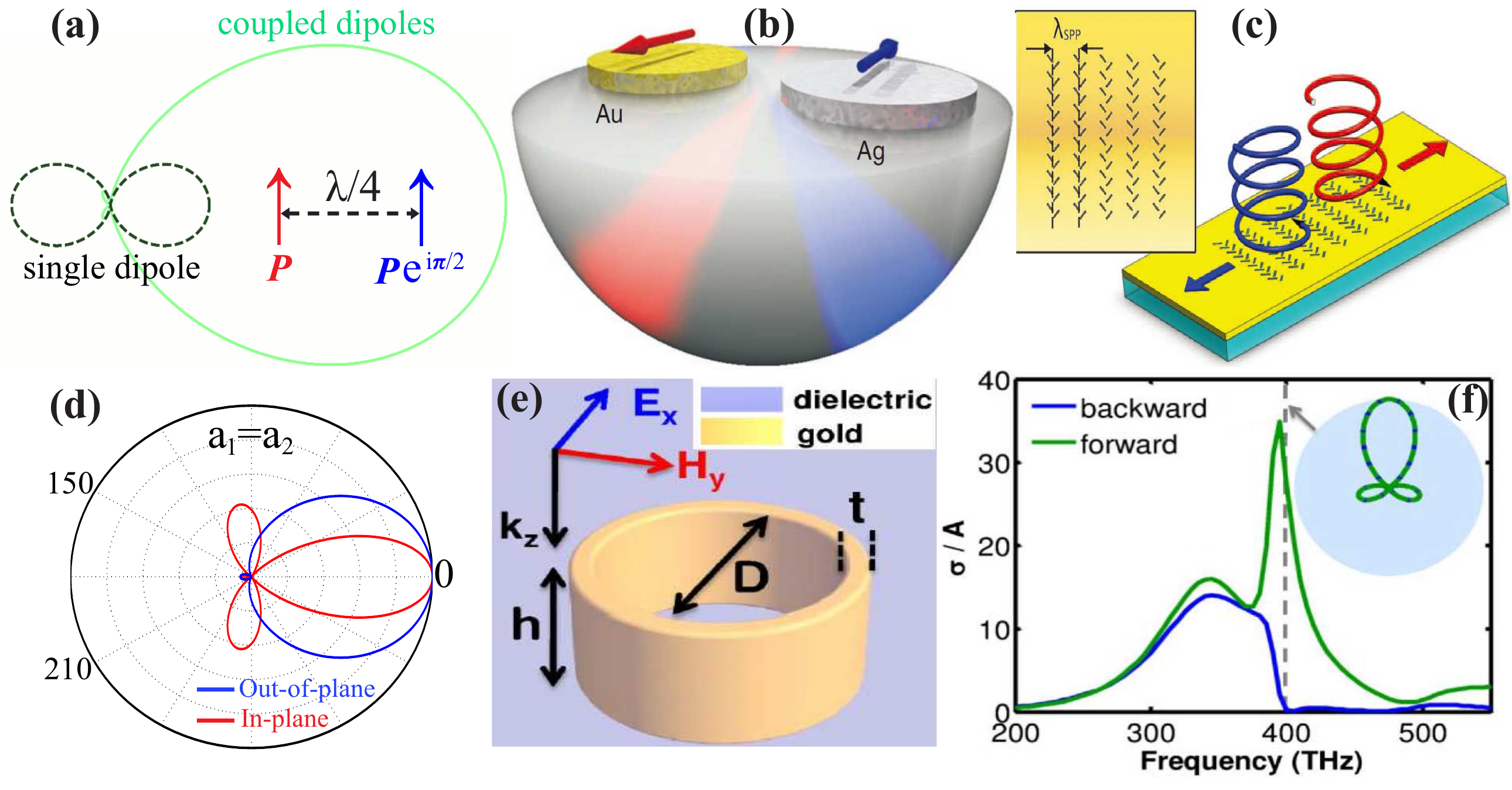}}\caption{\textbf{Interferences between electric multipoles}. (a) In-plane scattering patterns of a single dipole (dashed curve) and two separated  dipoles (solid curve). The two dipoles are of the same magnitude and they possess a $\pi/2$ phase difference. The distance between them is $\lambda/4$, where $\lambda$ is the effective wavelength of the interfering fields. (b) The interferences of two EDs supported by the metal discs can route light of different colours to different directions. (c) The flat metasurface consisting of perpendicular metal-bar pairs as unit cells can couple incident circularly polarized light into SPP waves propagating into  opposite directions which are helicity-dependent. To enhance the coupling efficiency, the horizontal periodicity is designed to be the effective wavelength of the SPP waves ($\lambda_{\rm SPP}$). (d) In-plane (red curve) and out-of-plane (blue curve) scattering patterns of the spherical particle with simultaneous dipole and quadrupole excitation ($a_1=a_2$).  (f) Perfect backward scattering elimination and forward scattering enhancement obtained by a metal ring (e) through interferences of the electric dipole and quadrupole modes. The inset in (f) shows the full in-plane angular scattering pattern. From: (a), Ref.~\cite{MIROSHNICHENKO_Science_polarization_2013}; (b), Ref.~\cite{SHEGAI_Nat.Commun._bimetallic_2011}; (c), Ref.~\cite{LIN_Science_polarizationcontrolled_2013}; (d), Ref.~\cite{liu_ultradirectional_2014}; (e) \& (f), Ref.~\cite{ALAEE_Opt.Lett._generalized_2015}. (Online version in colour.)}
\label{fig2}
\end{figure*}

In the field of nanophotonics, the investigations into interference between electric multipoles are most comprehensive, as magnetic and toroidal multipoles have attracted  attention only since the emergence  of the field of metamaterials~\cite{soukoulis_achievements_2011,ZHELUDEV_metamaterials_2012,kaelberer_toroidal_2010-1,PAPASIMAKIS_NatMater_electromagnetic_2016}. Before that, it had been long taken for granted that most nanostructures show dominantly electric responses with negligible magnetic responses, especially in the optical regime. As a result, there has been a significant amount of studies focusing on the electric responses only, where the analysis on the excitation and interferences of EDs plays a fundamental role~\cite{NOVOTNY_principles_2012,lukyanchuk_fano_2010-1,NOVOTNY_Nat.Photonics_antennas_2011,KILDISHEV_Science_planar_2013,CHEN_Rep.Prog.Phys._review_2016}. Radiation of an individual ED is symmetric, distributing energy equally into the two opposite directions perpendicular to the dipole orientation [see Fig.~\ref{fig2}(a)].  Some combinations of several EDs with engineered phases and amplitudes can render more freedom for the radiation pattern shaping, which is highly desired for many applications in nanophotonics such as sensing, nanoantennas and photovoltaic devices~\cite{NOVOTNY_principles_2012,KABASHIN_NatMater_plasmonic_2009,NOVOTNY_Nat.Photonics_antennas_2011,atwater_plasmonics_2010-1,liuwei_control_2014}. A well known example to break the radiation symmetry is the Yagi-Uda antenna~\cite{NOVOTNY_Nat.Photonics_antennas_2011}, where reflectors and directors are employed to route the radiation of the driven dipole to a preferred direction. In Fig.~\ref{fig2}(a), we show a much simpler while widely employed configuration of two coupled EDs with the $\pi/2$ phase difference and $\lambda/4$ displacement. A simple phase analysis reveals that the radiated fields of the two EDs interfere constructively and destructively toward the right and left, leading to a highly asymmetric scattering pattern compared to that of an individual ED [see Fig.~\ref{fig2}(a)]~\cite{MIROSHNICHENKO_Science_polarization_2013}. This is exactly the mechanism of the so-called anti-reflection coatings in optics~\cite{jackson_classical_1962,BOHREN_absorption_2008}.

A generalized version of the ED pair is that with the phase difference and displacement  properly tuned, then the constructive scattering enhancement can be observed at other radiation angles. An example of this is shown in Fig.~\ref{fig2}(b), where a pair of silver and gold disks can be viewed as two coupled EDs, and they can route light of different colours into different preferred directions~\cite{SHEGAI_Nat.Commun._bimetallic_2011}. This occurs due to the fact that for different wavelengths the phase accumulated through the optical path and the intrinsic phase difference (induced by different complex polarizabilities for two different material-dependent plasmon resonances~\cite{BOHREN_absorption_2008}) between two EDs of the metallic disks  vary, resulting in the functionality of flexible colour routing. A more recent exploitation of the concept of ED pairs is shown in Fig.~\ref{fig2}(c), where it is demonstrated that incident circularly polarized light of different handedness can be coupled to surface plasmon polariton (SPP) modes propagating into two opposite directions~\cite{LIN_Science_polarizationcontrolled_2013}.  The mechanism of such a helicity-dependent propagation is exactly the same as that shown in Fig.~\ref{fig2}(a), while here the circular polarization of incident waves provides two perpendicular metal-bar EDs in each unit cell an absolute $\pi/2$ phase difference. A different handedness can be employed to flip the sign of the phase difference and thus change the direction of SPP waves. Alternatively, such a helicity-dependent phase can be interpreted as a Pancharatnam-Berry phase~\cite{WILCZEK_geometric_1989,BOMZON_Opt.Lett._pancharatnamberry_2001,HUANG_LightSciAppl_helicity_2013}, based on which many other more sophisticated nanophotonics structures can be designed to enable more advanced applications~\cite{BLIOKH_Nat.Photonics_spinorbit_2015}.  Additionally, the  mechanism of coupled EDs can also be employed to shape radiation patterns of point emitters~\cite{BONOD_Phys.Rev.B_ultracompact_2010,ROLLY_Opt.Lett._crucial_2011} and to guide
the designs of various metasurfaces to realise different functionalities~\cite{KILDISHEV_Science_planar_2013,jahani_alldielectric_2016-1,Kuznetsov_Science_2016,CHEN_Rep.Prog.Phys._review_2016}.

The principles of interferences between EDs can be extended to higher-order multipoles, where the simplest case is the spatially-overlapping electric dipole and quadrupole modes. One fundamental platform for such an interference is the Mie scattering of spherical particles with incident plane waves~\cite{BOHREN_absorption_2008}.  A parity analysis shows that the electric multipoles of adjacent orders can interfere constructively and destructively in the forward and backward directions, respectively~\cite{liu_ultradirectional_2014}.  In Fig.~\ref{fig2}(d) we show the scattering of the interfering electric dipole and quadrupole (with equal scattering coefficients $a_1=a_2$), where a highly asymmetric scattering pattern can be obtained~\cite{liu_ultradirectional_2014,lukyanchuk_fano_2010-1}.
Such principle can certainly be applicable to other relatively irregular structures, such as split ring resonators~\cite{HANCU_NanoLett._multipolar_2014} and V-antennas~\cite{VERCRUYSSE_ACSNano_directional_2014,LI_NanoLett._alldielectric_2016}

To fully eliminate the scattering at the backward direction [in Fig.~\ref{fig2}(d), the backward scattering is suppressed but not fully eliminated], the scattering coefficients should satisfy the condition of $3a_1=5a_2$, which is challenging to meet as it involves the perfect matching for both real and imaginary parts of the scattering coefficients.  This has recently been achieved within a metal ring [see Fig.~\ref{fig2}(e)], where the backward scattering can be fully eliminated while the forward scattering is enhanced [see Fig.~\ref{fig2}(f) and the inset] with properly tuned multipole amplitude and phase ~\cite{ALAEE_Opt.Lett._generalized_2015}.  We notice that here we confine our discussion to  the case of spatially overlapping two multipoles of low orders, and a more exhaustive employment of electric multipolar interferences can involve more spatially separated multipoles of higher orders~\cite{lukyanchuk_fano_2010-1,ZHANG_Phys.Rev.Lett._plasmoninduced_2008}, which can provide much more flexibilities for nanoscale light control.

\section{Magnetic multipoles}
\label{mm}

\begin{figure*}
\centerline{\includegraphics[width=11cm]{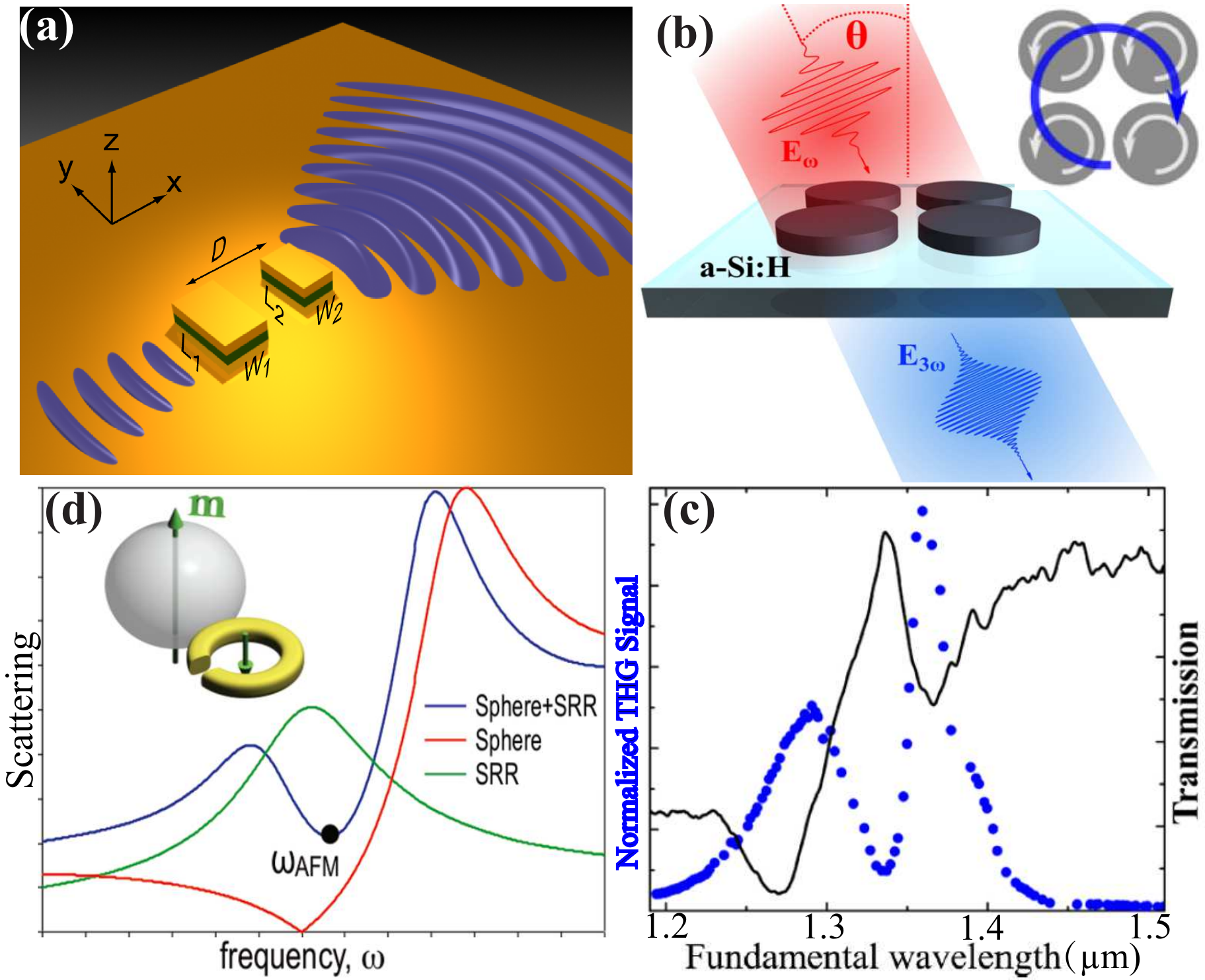}}\caption{\textbf{Interferences between magnetic multipoles}. (a) Two interfering metal-dielectric resonators, which can be viewed as two MDs, can couple the incident light into SPP waves propagating mainly along one preferred direction. (b) All-dielectric  quadrumer made of a-Si:H disks supports both narrow MDs by each individual disk and a broad collective MD (see inset), which can interfere with one another to produce the magnetic Fano resonance, and (c) the corresponding spectra for both fundamental wave and normalised third-harmonic generation. At the Fano dip close to the fundamental wavelength of $1.35~\mu$m,  there is significant near-field enhancement within the quadrumer, which leads to significant enhancement for third-harmonic generation. (d)  Scattering intensity spectra for a single SRR, a single dielectric sphere, and coupled SRR and dielectric sphere. The strong coupling among the individual resonances can render the two MDs out of phase (see inset), leading to the effective antiferromagnetic structures. From: (a), Ref.~\cite{LIU_NanoLett._compact_2012}; (b) \& (c), Ref.~\cite{SHOROKHOV_NanoLett._multifold_2016}; (d), Ref.~\cite{MIROSHNICHENKO_ACSNano_optically_2012}. (Online version in colour.)}
\label{fig3}
\end{figure*}

The symmetry of Maxwell's equations explains that the electric and magnetic multipoles share some common properties: for example, EDs and MDs oriented along the same direction would have identical radiation patterns~\cite{NOVOTNY_principles_2012,jackson_classical_1962,BOHREN_absorption_2008}.  Generally speaking, all the principles discussed in the Sec.~\ref{ee} above for the electric multipolar interferences can be directly mapped to the case of magnetic multipoles. For example, the mechanism revealed in Fig.~\ref{fig2}(a) applies equally to the case of two coupled MDs, when highly asymmetric scattering patterns can be also obtained. Nevertheless, the magnetic responses of nanostructures attracted a special attention only recently after the emergence of the field of metamaterials, where the concept of optically-induced magnetic response is playing a central role~\cite{soukoulis_achievements_2011,ZHELUDEV_metamaterials_2012}. Many types of effective magnetic multipoles have been found to exist not only in traditional plasmonic nanostructures, but also in all-dielectric structures~\cite{KRASNOK_Phys.-Uspekhi_optical_2013,liuwei_control_2014,jahani_alldielectric_2016-1,Kuznetsov_Science_2016}.  This fosters many studies focusing on extending light manipulation principles based on the interferences of electric multipoles to the case of magnetic multipoles and their combinations. As an example, in Fig.~\ref{fig3}(a) we show the interferences of two coupled metal-dielectric resonators which can be viewed as a pair of MDs~\cite{LIU_NanoLett._compact_2012}. Similar to the case shown in Fig.~\ref{fig2}(b), the two resonators are of different sizes and thus of different complex polarizabilities, which defines the intrinsic phase difference between them. This polarizability-induced phase, together with the phase accumulated along the gap between two resonators, can render the incident wave coupled to the SPP modes propagating along one preferred direction.

Another noticeable example of the direct mapping from the electric to magnetic multipoles is the magnetic Fano resonances observed for all-dielectric nanoparticle clusters or nanoparticle arrays~\cite{EVLYUKHIN_Phys.Rev.B_optical_2010,miroshnichenko_fano_2012,HOPKINS_ACSPhotonics_interplay_2015,SHOROKHOV_NanoLett._multifold_2016}, being somewhat analogous to the electric Fano resonance observed for metalic particle counterparts~\cite{miroshnichenko_fano_2012,lukyanchuk_fano_2010-1,liu_polarizationindependent_2012}. A recently studied nanodisk quadrumers
fabricated of hydrogenated amorphous silicon (a-Si:H) is shown in Fig.~\ref{fig3}(b), and the spectra for both fundamental wave and normalized third-harmonic generation are shown in Fig.~\ref{fig3}(c) (incident angle $\theta=\pi/4$)~\cite{SHOROKHOV_NanoLett._multifold_2016}. In brief, an obliquely incident wave can excite simultaneously two kinds of MDs:  a narrow MD of each individual dielectric nanodisk and collective broad MD of the whole quadrumer structure~\cite{HOPKINS_ACSPhotonics_interplay_2015} [see the inset of Fig.~\ref{fig3}(b)]. The narrow and broad MDs  interfere with one another to produce an asymmetric Fano resonance, which is directly driven by optically-induced magnetism~\cite{lukyanchuk_fano_2010-1}. This was verified in experiment by measuring the transmission shown in Fig.~\ref{fig3}(c). At the Fano dip (close to the fundamental wavelength of $1.35~\mu$m), the fields within the quadrumer can grow significantly, and thus at the Fano resonance the third-harmonic generation can be enhanced dramatically [see Fig.~\ref{fig3}(c)].

Interferences of magnetic multipoles can also produce many other exotic electromagnetic effects, an outstanding example of which is the recently proposed effective optical antiferromagnetism~\cite{MIROSHNICHENKO_ACSNano_optically_2012}. Figure~\ref{fig3}(d) shows the scattering intensity of the antiferromagnetic metamaterials based on the compact MD pairs with a $\pi$ phase difference supported respectively by a split-ring resonator (SRR) and dielectric sphere~\cite{MIROSHNICHENKO_ACSNano_optically_2012}.  Basically, the strong coupling between multiple resonances can render the two MDs out of phase even though they are very close to each other [see the inset in Fig.~\ref{fig3}(d)], leading to an effectively overall staggered (or antiferromagnetic) ordering of the optically-induced magnetic moments.

It is worth mentioning that, although in this section we have discussed the interferences of magnetic multipoles, this does not mean that there are no contributions of electric multipoles. Actually in the configuration shown in Fig.~\ref{fig3}(d) the role played by EDs of the dielectric sphere is crucial, without which the out-of-phase MD pairs and thus the effective antiferromagnetism is not accessible. Moreover, not only the collective MD but also individual MDs shown in Figs.~\ref{fig3}(a,b) can be interpreted alternatively as a combination of out-of-phase EDs. A noticeable simple example of this is that within strongly coupled metal bars (which if isolated can be interpreted as an ED), MDs can be effectively excited~\cite{SVIRKO_Appl.Phys.Lett._layered_2001,SHALAEV_Opt.Lett._negative_2005}. For the configurations shown in Fig.~\ref{fig1}(b,c), as the metallic nanoparticles are well separated and thus not strongly coupled to each other,  the mechanism of two-ED interference shown Fig.~\ref{fig1}(a) is still applicable.
In principle however, all photonic nonmagnetic structures can be decomposed into many interacting discrete small EDs~\cite{DRAINE_JOSAA_discretedipole_1994}, and the electromagnetic effects can be explained by the interferences of those EDs only. Nevertheless, under most circumstances the adoption of the concept of magnetic multipoles and higher-order electric multipoles can render deeper physical insights and bring great convenience for various structure designs and applications. We notice that here we restrict our discussions to MDs only. Similar to the electric multipoles discussed above in Sec.~\ref{ee}, such consideration can be naturally extended to higher-order magnetic multipoles, the excitation of which are highly accessible for various structures, especially those high permittivity dielectric particles (See Refs.~\cite{liuwei_control_2014,jahani_alldielectric_2016-1,Kuznetsov_Science_2016,CHEN_Rep.Prog.Phys._review_2016} and references therein). It is actually easy to obtain the magnetic counterparts of what is shown in Figs.~\ref{fig2}(d-e) and many other electric multipolar interference effects.

\begin{figure*}
\centerline{\includegraphics[width=14cm]{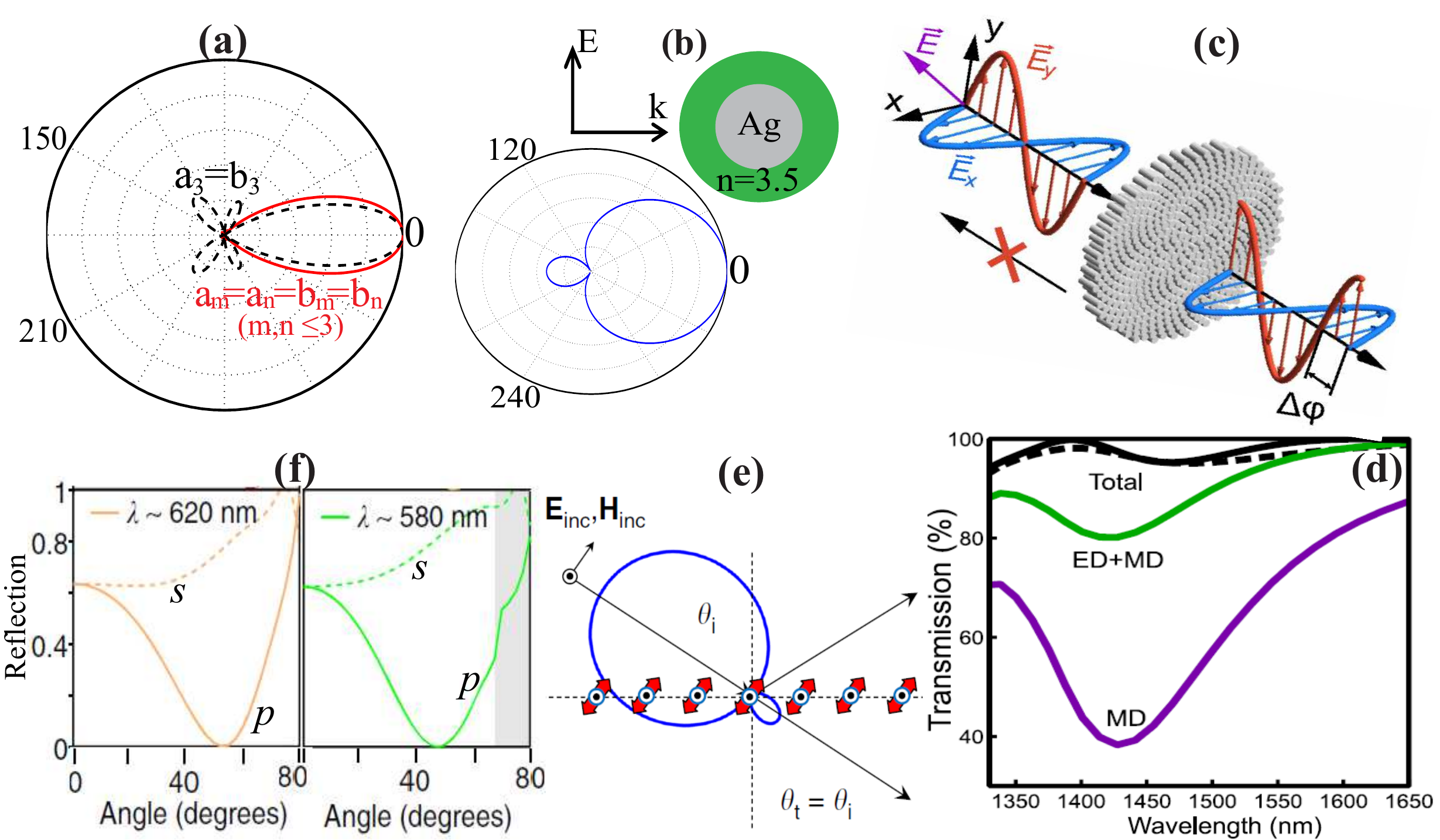}}\caption{ \textbf{Interferences between electric and magnetic multipoles}. (a) Mie scattering patterns of spherical particles  by incident plane waves. Dashed curve: overlapping electric and magnetic octopoles only with the same magnitude ($a_3=b_3$); Solid curve: overlapping electric and magnetic dipoles, quadrupoles and octopoles with the same corresponding scattering coefficients [$a_m=a_n=b_m=b_n~(m,~n\leq3$)]. (b) Mie scattering pattern of a core-shell nanowire by an incident TM plane wave. The nanorods support overlapping ED and MD, and the magnitude of ED is twice of that of MD, leading to scattering elimination at $\theta=2/3\pi$ and $4/3\pi$. (c) Silicon nanorod metasurface with suppressed reflection and flexible transmission phase control, and (d) the transmission spectra (with also the contribution from MD, and the joint contribution of ED and MD). (e) An array of meta-atoms made of perpendicular ED-MD dipole-pairs of which the scattering suppression angle can be controlled. (f) The wavelength-dependent reflection-incident angle spectra [for both $p$ (solid) and $s$ (dashed) polarized reflected waves]  of the metasurface consisting of a two dimensional square lattice of silicon spheres. The incident wave is $p$-polarized. From: (a), Ref.~\cite{liu_ultradirectional_2014}; (b), Ref.~\cite{liu_scattering_2013-1}; (c) \& (d), Ref.~\cite{KRUK_APLPhotonics_invited_2016}; (e) \& (f), Ref.~\cite{PANIAGUA-DOMINGUEZ_NatCommun_generalized_2016-1}. (Online version in colour.)}
\label{fig4}
\end{figure*}

\section{Electric and magnetic multipoles}
\label{em}

In nanophotonics, the incorporation of magnetic responses provides an extra degree of freedom for the efficient light control through interfering electric and magnetic multipoles~\cite{KRASNOK_Phys.-Uspekhi_optical_2013,liuwei_control_2014,jahani_alldielectric_2016-1,Kuznetsov_Science_2016,CHEN_Rep.Prog.Phys._review_2016}. One of the most noticeable examples is the recent demonstration of simultaneous forward scattering enhancement and backward scattering suppression based on interferences of EDs and MDs (the so-called Kerker's  scattering)~\cite{KERKER_J.Opt.Soc.Am._electromagnetic_1983,liuwei_control_2014}.  The simplest case of dipolar Kerker scattering and its applications in various nanostructures including metasurfaces have been discussed extensively in several reviews~\cite{KRASNOK_Phys.-Uspekhi_optical_2013,liuwei_control_2014,jahani_alldielectric_2016-1,Kuznetsov_Science_2016,CHEN_Rep.Prog.Phys._review_2016}, and recently such principle has been applied for strongly coupled particle clusters~\cite{YAO_ACSPhotonics_controlling_2016} and antennas made of switchable phase-change materials~\cite{ALAEE_Opt.Lett._phasechange_2016}.
Here focus on two generalizations of the dipolar Kerker scattering: the scattering based on interference of electric and magnetic multipoles of higher orders, and electric and magnetic multipoles of different magnitudes and phases.

For the seminal problem of Mie scattering by spherical particles, it is shown that electric and magnetic multipoles of the same order show opposite parities with respect to $\cos\theta$ ($\theta$ is the scattering angle with respect to the forward direction)~\cite{liu_ultradirectional_2014}. This means that simultaneous forward scattering enhancement and backward scattering suppression can be achieved for not only interfering EDs and MDs, but also for any higher order multipoles where the forward scattering directionality can be further enhanced~\cite{liu_ultradirectional_2014}.  In Fig.~\ref{fig4}(a) we show the scattering pattern of interfering electric and magnetic octopoles only (dashed curve, $a_3=b_3$, other multipoles are negligible). Compared to the dipolar case~\cite{liu_broadband_2012,liu_ultradirectional_2014}, the enhanced forward scattering is more directional, but with several side scattering lobes.  To suppress the side scattering lobes, the interferences between the same type of multipoles of  different orders [as discussed in Section~\ref{ee} and shown in Fig.~\ref{fig2}(d)] can be employed. We further show in Fig.~\ref{fig4}(a) the scattering pattern produced by interfering electric and magnetic dipoles, quadrupoles and octopoles with the same corresponding scattering coefficients [solid curve, $a_m=a_n=b_m=b_n~(m, n\leq3$)]. It is clear that the side scattering has been eliminated without compromising much of the forward scattering directionality. Such an approach to optimize the scattering directionality based on generalized Kerker scattering has been realized for plasmonic core-shell particles, for gap SPP resonators on substrate~\cite{PORS_Opt.Express_unidirectional_2015}, for coupled dipolar emitter and dielectric-magnetic particle systems~\cite{ROLLY_Sci.Rep._controllable_2013,KRASNOK_Nanoscale_superdirective_2014,RUSAK_Appl.Phys.Lett._hybrid_2014}, and for generated nonlinear harmonic waves ~\cite{SMIRNOVA_Optica_Multipolar_2016,WANG_ACSPhotonics_multipolar_2016,CARLETTI_ACSPhotonics_shaping_2016,SMIRNOVA_ACSPhotonics_multipolar_2016}.

The Kerker scattering can also be generalized along another direction to electric and magnetic multipoles of different amplitudes and phases, where then the scattering elimination can be achieved in other  angles required~\cite{liu_scattering_2013-1,PANIAGUA-DOMINGUEZ_NatCommun_generalized_2016-1}.  In Fig.~\ref{fig4}(b) we show the two dimensional transverse-magnetic (TM) wave  scattering by a metal core-dielectric shell nanowire, where the EDs and MDs can be tuned to spectrally overlap~\cite{liu_scattering_2013-1}. As the ED corresponds to two degenerate channels, the magnitude of ED is twice of that of the MD, rendering the scattering elimination at scattering angles of $\theta=2/3\pi$ and $4/3\pi$ rather than the backward direction. Such interference induced  control of scattering elimination angle can be also realized in three dimensional spheres through tuning the EDs and MDs supported~\cite{PANIAGUA-DOMINGUEZ_NatCommun_generalized_2016-1}.

When applied to metasurfaces~\cite{KILDISHEV_Science_planar_2013,jahani_alldielectric_2016-1,Kuznetsov_Science_2016,CHEN_Rep.Prog.Phys._review_2016}, the two versions of generalized Kerker scattering discussed above can provide new stimuli and physical insights~\cite{YANG_NanoLett._dielectric_2014,PANIAGUA-DOMINGUEZ_NatCommun_generalized_2016-1,KRUK_APLPhotonics_invited_2016,PROUST_ACSNano_alldielectric_2016}. In Fig.~\ref{fig4}(c) we show the metasurface consisting of silicon nanorods, which support not only EDs and MDs, but also higher order multipoles~\cite{KRUK_APLPhotonics_invited_2016}.  The total transmission spectra of such a metasurface is shown in Fig.~\ref{fig4}(d), where the  contributions from MD and a combination of ED and MD are also shown. It is clear that the excitation of higher order multipoles besides dipoles contributes to the outstanding feature of broadband near-unity transmission of such a metasurface, the mechanism of which is exactly the generalized Kerker scattering discussed above [see Fig.~\ref{fig4}(a)]. Moreover, the coexistence of higher order electric and magnetic multipoles also provides much more flexibilities for phase control of the transmitted wave, enabling the functionality of highly efficient polarization control~\cite{KRUK_APLPhotonics_invited_2016}.

The other case of generalized Kerker scattering can be employed to metasurfaces based on which the concept of Brewster angle can be broadened~\cite{PANIAGUA-DOMINGUEZ_NatCommun_generalized_2016-1}.  As shown in Fig.~\ref{fig4}(e), if we arrange the meta-atom consisting of a pair of perpendicular ED and MD (the angle of scattering elimination can be controlled through tuning the dipolar magnitude and phase) into arrays on a surface, the angle of zero scattering from such an array can be tuned. In Fig.~\ref{fig4}(f) we show the wavelength-dependent reflection-incident angle spectra (for both $p$ and $s$ polarized reflected waves; the incident wave is $p$-polarized with the electric field on the incident plane) for a metasurface made of two dimensional square lattice of silicon spheres.  For each sphere at different wavelengths, the ratios between the magnitudes of EDs and MDs vary, leading to scattering suppression at different angles~\cite{liu_scattering_2013-1,PANIAGUA-DOMINGUEZ_NatCommun_generalized_2016-1}.  For a square lattice of such spheres, this would result in the wavelength-dependent Brewster angles, as verified in Fig.~\ref{fig4}(f). Moreover, the wavelength-dependent Brewster angle can be also observed for $s$-polarized incident waves~\cite{PANIAGUA-DOMINGUEZ_NatCommun_generalized_2016-1}, which is unaccessible at conventional interfaces where there is only dominant ED excitation. Here we discuss only the scattering suppression at different angles based on interferences between ED and MD, which can certainly be extended to higher order multipoles.

\section{Electric and toroidal multipoles}
\label{et}

Though the dynamic toroidal multipoles play an essential role in the expansions for arbitrary charge-current distributions~\cite{jackson_classical_1962,PAPASIMAKIS_NatMater_electromagnetic_2016}, they had not attracted much attention until recent demonstration of TDs in the engineered metamaterials~\cite{kaelberer_toroidal_2010-1}. This is largely due to the fact that in the far field toroidal multipoles are identical to their electric multipolar counterparts~\cite{PAPASIMAKIS_NatMater_electromagnetic_2016}. At the same time, conventional multipole expansions are usually confined to charge-current distributions that are far smaller than the effective wavelength, where the contributions of toroidal multipoles are negligible~\cite{jackson_classical_1962,liu_efficient_2015}.  Though the isolated excitation of dynamic toroidal multipoles would require careful structure engineering to match the special corresponding near-field current distributions~\cite{kaelberer_toroidal_2010-1,PAPASIMAKIS_NatMater_electromagnetic_2016}, it is also shown recently that even simple homogeneous dielectric particles can support TDs~\cite{MIROSHNICHENKO_NatCommun_nonradiating_2015,liu_toroidal_2015,liu_invisible_2015,liu_efficient_2015}, though they are co-excited with the electric and magnetic multipoles. Roughly speaking, to effectively support dynamic TDs, the excitation of out-of-phase MDs are required.  Even for the fundamental homogeneous spherical and cylindrical structures, such out-of-phase MDs can be excited for relatively large particle sizes, leading to the effective formation of dynamic TDs~\cite{MIROSHNICHENKO_NatCommun_nonradiating_2015,liu_toroidal_2015,liu_invisible_2015,liu_efficient_2015}. The excitation of TDs within composite structures has been discussed in detail in the Ref.~\cite{PAPASIMAKIS_NatMater_electromagnetic_2016}, and here we focus on the excitation of TDs within individual nanoparticles and their interferences with the electric counterparts (EDs).

\begin{figure*}
\centerline{\includegraphics[width=12cm]{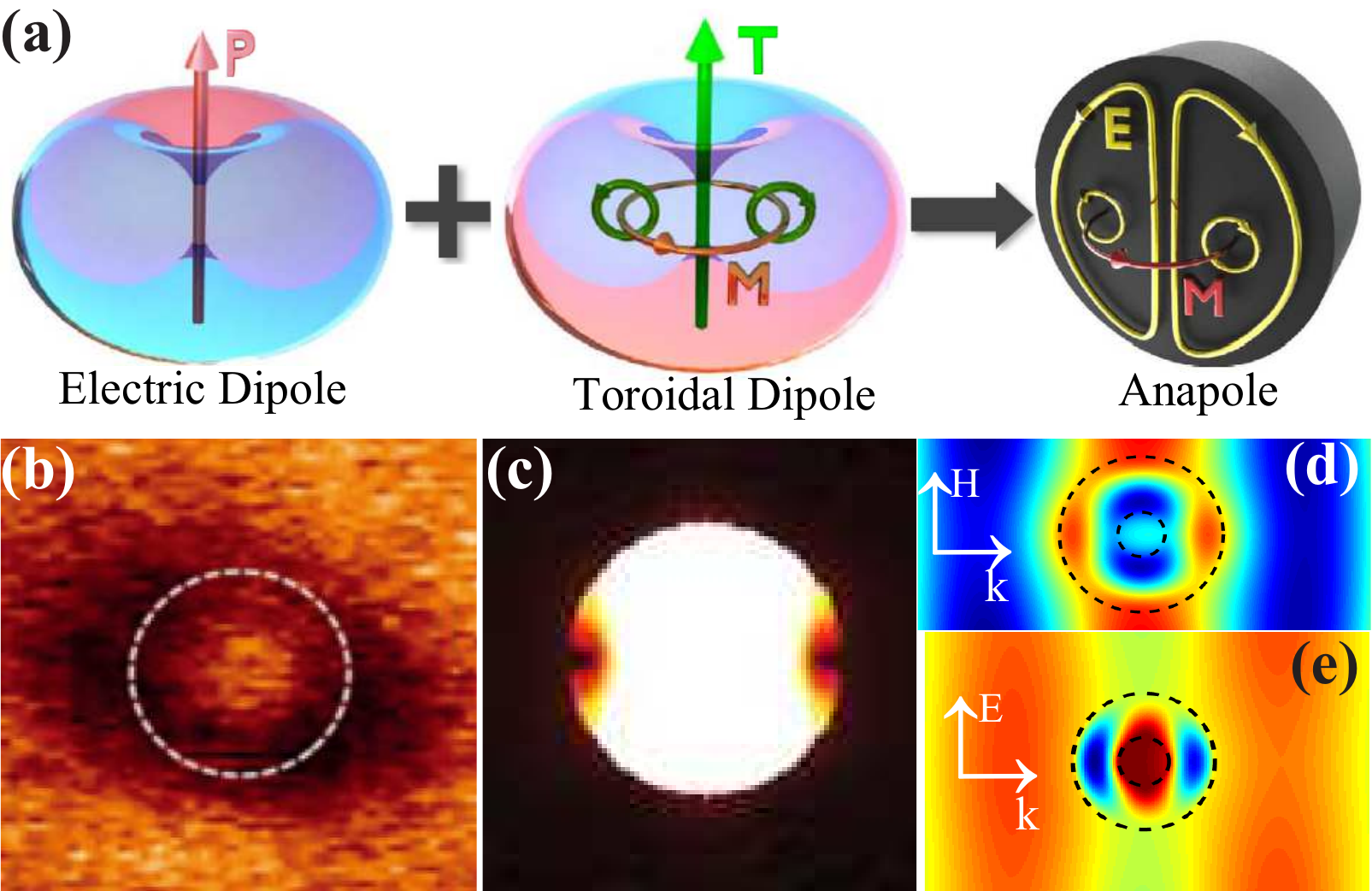}}\caption{ \textbf{Interferences between electric and toroidal multipoles}. (a) The destructive interference of an ED (left) and a TD (middle) leads to the formation of a nonradiating anapole mode.  The near-field distributions for the anapole modes excited within: (b) silicon nanodisk with plane wave incidence; (c) all-dielectric nanosphere with  two counter propagating radially polarized beams of the same intensity but out of phase (the incident wave is subtracted from the field distribution plot); (d) \& (e) metal-dielectric core-shell nanowires of both polarized incident plane waves. From: (a) \& (b), Ref.~\cite{MIROSHNICHENKO_NatCommun_nonradiating_2015}; (c), Ref.~\cite{WEI_Optica_excitation_2016}; (d) \& (e), Ref.~\cite{liu_invisible_2015}. (Online version in colour.)}
\label{fig5}
\end{figure*}

Since the EDs and TDs have the identical scattering patterns, it means that when co-excited and spatially overlapped with the same scattering magnitude but out of phase,  they can cancel the scattering of each other in the far field, appearing to be invisible~\cite{MIROSHNICHENKO_NatCommun_nonradiating_2015,PAPASIMAKIS_NatMater_electromagnetic_2016} [see   Fig.~\ref{fig5}(a)]. This is the basic mechanism of the recently discovered non-radiating anapoles~\cite{MIROSHNICHENKO_NatCommun_nonradiating_2015}.  To observe a pure non-radiating anapole excitation, besides proper ED and TD excitation, the suppression of other multipoles is also required. This is not possible for homogeneous spheres or cylinders with incident plane waves as the MDs and other quardupolar excitations at the ED-TD scattering cancelling point is noneligible~\cite{MIROSHNICHENKO_NatCommun_nonradiating_2015,liu_toroidal_2015,liu_invisible_2015,liu_efficient_2015}.  The anapole excitation has been firstly experimentally realised within engineered composite metallic metamaterials in the microwave regime, where the excitation of other multipoles are negligible~\cite{fedotov_resonant_2013}. The pure anapole excitation can be also achieved with other unwanted  multipoles suppressed within the following individual nanoparticles in the optical regime: (i) All-dielectric nanodisks with incident plane waves~\cite{MIROSHNICHENKO_NatCommun_nonradiating_2015} [as is shown in Fig.~\ref{fig5}(b) of the experimental near-field enhancement close to the silicon nanodisk at the anapole point]; (ii) All-dielectric spheres with engineered incident waves (two counter propagating
radially polarized beams with the same intensity but out of phase)~\cite{WEI_Optica_excitation_2016}. The electric field distribution (incident wave subtracted) is shown in Fig.~\ref{fig5}(c) at the anapole point; (iii) Core-shell metal-dielectric nanospheres or nanorods with incident plane waves~\cite{liu_toroidal_2015,liu_invisible_2015}. Figure~\ref{fig5}(d) \& (e) shows the electric field distributions of the core-shell cylinders for both polarizations of the incident plane waves at the anapole point ~\cite{liu_invisible_2015}. For both cases the incident wave has experienced almost no perturbations, verifying the non-radiative nature of the anapole mode excited. It is recently also shown that for homogeneous spherical particles with incident plane waves, the radial anisotropy can be employed for pure anapole excitation~\cite{LIU_J.Nanomater._elusive_2015}.  The efficient excitation of non-radiating anapoles can play a significant role for improving nonlinear conversion efficiency and enhancing absorption in photonic nanostructures~\cite{GRINBLAT_NanoLett._enhanced_2016,WANG_Opt.Express_engineering_2016}. It is worth mentioning that here we confine our discussions to anapoles induced by interferences between ED and TD only, and it is natural to extend the discussions to higher-order anapoles, which are induced by the scattering cancellation of a pair of electric and toroidal multipoles of higher orders~\cite{PAPASIMAKIS_NatMater_electromagnetic_2016}.


\section{Concluding remarks and outlook}
\label{co}

As a source of electromagnetic fields, the excited electric charges and currents can generate the multipole modes of different types and orders, and their interference define many important electromagnetic effects at the nanoscale. Here we have discussed the multipolar interference effects in resonant metallic, metal-dielectric, and dielectric nanophotonics structures. More specifically,  we have presented a coherent view into interference effects involving electric, magnetic and toroidal multipoles, and have shown that such interferences lead directly to many exotic nanophotonic effects, such as ultra-directional propagation of light, optical antiferromagnetism, generalized Kerker scattering and Brewster angle, as well as recently discovered nonradiating optical anapoles. A complete employment of interferences between all sorts of multipoles of different types and orders relying on both amplitude and phase engineering can provide much more flexibilities for nanoscale manipulation of free-space light propagation and light-matter interaction, which might incubate and stimulate new ideas for fundamental researches and many applications in nanophotonics.

As marked by the red double arrow and circle in Fig.~\ref{fig1}(a) above, the interferences between toroidal multipoles and those between magnetic and toroidal multipoles have not been well studied in the existing literatures. Nevertheless, since electric and toroidal multipoles of the same order have identical scattering patterns, it is natural to expect that the basic principles discussed in Sec.~\ref{ee} for  electric multipoles can be mapped directly to toroidal multipoles, meaning that interferences of toroidal multipoles can also be employed to produce highly directional light propagation and to realize the toroidal Fano resonances. The mechanisms discussed in Sec.~\ref{em} for interferences between electric and magnetic multipoles can also be applied directly to magnetic and toroidal multipoles, indicating that the generalized Kerker scattering and Brewster angle are also accessible based on interferences between magnetic and toroidal multipoles. We notice that even for the well-studied cases discussed in Secs.~\ref{ee}-\ref{et}, most studies are confined to interferences of low-order multipoles (up to quadrupoles only). Moreover, the multipolar interference can also result in other exotic phenomena including chiral and birefringent effects in nanophotonics~\cite{ZHANG_Phys.Rev.Lett._negative_2009,PLUM_Phys.Rev.Lett._metamaterials_2009,PAPASIMAKIS_Phys.Rev.Lett._gyrotropy_2009,soukoulis_achievements_2011}, and actually the optical chirality can involve the interacting multipoles from all three families~\cite{PAPASIMAKIS_Phys.Rev.Lett._gyrotropy_2009,PAPASIMAKIS_NatMater_electromagnetic_2016}. Basically, there are still many intriguing topics to explore in this research field, and a thorough investigation and exploitation of interferences between multipoles of higher orders and of all three major families would give more impetus to the progress of nanophotonics. Additionally, the analysis based on multipole expansions and interferences can also be extended to newly emerging fields that hybridize quickly with the subject of nanophotonics, such as quantum photonics~\cite{TAME_NatPhys_quantum_2013,SCHRODER_ArXiv160305339Cond-MatPhysicsphysicsPhysicsquant-Ph_review_2016}, two-dimensional flatland photonics~\cite{xia_twodimensional_2014}, and topological photonics~\cite{LU_NatPhoton_topological_2014}. The analysis of multipolar interference effects in those interdisciplinary fields may help to gain deeper
physical insights into many electromagnetic effects, which might foster more fundamental research and stimulate advanced applications.





We thank  A.E. Miroshnichenko, C. Rockstuhl, and J. Zhang for useful comments and suggestions. This work was supported by the National Natural Science Foundation of China (Grant number: $11404403$), the Basic Research Scheme of College of Optoelectronic Science and Engineering, National University of Defence Technology (China), and several grants of the Australian Research Council.




\end{document}